\documentclass[12pt]{article}

\begin{document}
\renewcommand{\thefootnote}{\fnsymbol{footnote}}

\begin{titlepage}
\hfill{hep-th/0208074}
\vspace{15mm}
\baselineskip 8mm

\begin{center}
{\LARGE \bf ${\cal N}$=(4,4) Type IIA String Theory \\
on PP-Wave Background }
\end{center}
\baselineskip 6mm
\vspace{10mm}
\begin{center}
Seungjoon Hyun$^a$\footnote{\tt hyun@phya.yonsei.ac.kr} and
Hyeonjoon Shin$^b$\footnote{\tt hshin@newton.skku.ac.kr} \\[5mm]
{\it
$^a$Institute of Physics and Applied Physics, Yonsei University,
Seoul 120-749, Korea \\
$^b$ BK 21 Physics Research Division and Institute of Basic Science \\
Sungkyunkwan University, Suwon 440-746, Korea}
\end{center}

\thispagestyle{empty}


\vfill

\begin{center}
{\bf Abstract}
\end{center}
\noindent 
We construct IIA GS superstring action on the ten-dimensional pp-wave
background, which arises as the compactification of eleven-dimensional
pp-wave geometry along the isometry direction. The background geometry
has 24 Killing spinors and among them, 16 components correspond to the
non-linearly realized kinematical supersymmetry in the string
action. The remaining eight components are linearly realized and shown
to be independent of $x^+$ coordinate, which is identified with the
world-sheet time coordinate of the string action in the light-cone
gauge. The resultant dynamical $\cal{N}$=(4,4) supersymmetry is
investigated, which is shown to be consistent with the field contents
of the action containing two free massive supermultiplets.
\vspace{20mm}
\end{titlepage}

\baselineskip 6.5mm
\renewcommand{\thefootnote}{\arabic{footnote}}
\setcounter{footnote}{0}


\section{Introduction}
In most cases, M theory is much easier to handle when it is
compactified along the small circle, in which the theory becomes the
weakly-coupled IIA superstring theory. In this case we can use full
power of perturbative string theory to probe the model. It may be true
even in the M theory on non-trivial background such as pp-wave
geometry \cite{kow194}-\cite{bla081}. The pp-wave geometry for IIB
string theory is maximally supersymmetric\cite{bla242}. Furthermore,
the IIB superstring theory on the pp-wave background reduces to free
massive theory in the light-cone gauge \cite{met044} and thus many
techniques in string theory on the flat Minkowski background can be
adopted.  The theory is used to probe the stringy nature of
four-dimensional ${\cal N}=4$ super Yang-Mills theory in
\cite{ber021}.

On the other hand, even though M theory on the eleven-dimensional
pp-wave is also maximally supersymmetric, it is not easy to probe
various aspects on the theory due to the lack of powerful tool as much
as string theory. Matrix model on pp-wave background
\cite{ber021,das185} has interesting property such as the removal of
flat directions due to the presence of mass terms. Subsequently,
various aspects on matrix theory in pp-wave background has been
investigated \cite{hyu090}-\cite{sug070}.  However it has
time-dependent supersymmetry which does not commute with the
Hamiltonian.  The implication of the time-dependent supersymmetry is
not clear. And still it may be desirable to study the model in the
context of ten-dimensional string theory, more or less like the IIB
string theory on pp-wave background.

In this paper we consider the Type IIA superstring theory on the
following ten-dimensional pp-wave background:
\begin{eqnarray}
& & ds^2 = - 2 dx^+ dx^-
    - A(x^I) (dx^+)^2  + \sum^8_{I=1} (dx^I)^2~,
                                      \nonumber \\
& & F_{+123} = \mu~,\ \ \  F_{+4} = -\frac{\mu}{3}~,
\label{pp-wave}
\end{eqnarray}
where
\begin{equation}
A(x^I) = \sum^4_{i=1} \frac{\mu^2}{9} (x^i)^2
            +\sum^8_{i'=5} \frac{\mu^2}{36} (x^{i'})^2~.
\end{equation}
As shown in the appendix, this geometry comes from the
compactification of the eleven-dimensional pp-wave geometry along the
isometry direction\cite{mic140}.\footnote{This is different pp-wave
  geometry from the one used by Alishahiha et.al. in \cite{ali037}.
  Their geometry comes from the T-duality of IIB pp-wave.  See also
  Refs. \cite{ben195,mic204} for similar construction.}  It has 24
supersymmetry, thus one may fear that the theory is not so simple
compared to the case of maximally supersymmetric IIB string theory on
pp-wave geometry.  As we will show, 16 supersymmetry in target space
is non-linearly realized on the string worldsheet theory. On the other
hand, eight supersymmetry is linearly realized.  This eight,
so-called, dynamical supersymmetry, which is half to the maximally
symmetric cases, turns out to be time independent. This means that
each half of bosonic fields have the same mass as each half of
fermionic fields, forming a supermultiplet.  Thus the theory is
described by the two-dimensional ${\cal N}$=(4,4) supersymmetric
theory with two massive supermultiplets.  The light-cone action of the
IIA superstring on the above pp-wave geometry turns out to have only
quadratic terms, the same as the case of the IIB superstring theory on
the pp-wave, thus the theory seems to be almost as simple as the IIB
case.

In section 2, we show that the pp-wave geometry admits 24 Killing
spinors.  In particular, we find the explicit form of the Killing
spinors for later use.  In section 3, we construct the ligh-cone
superstring action on the above pp-wave background starting from the
eleven dimensional supermembrane action on $AdS_4 \times S^7$ in the
pp-wave limit.  The action becomes the one of two massive
supermultiplets of (4,4) supersymmetry.  In section 4, we identify the
kinematical and dynamical supersymmetry on the worldsheet action.  In
particular, we give the transformation rules for the ${\cal N}= (4,4)$
worldsheet supersymmetry.  In section 5, we draw some conclusions. In
the appendix, we describe the compactification of the
eleven-dimensional pp-wave which gives rise to the IIA pp-wave
geometry and give some useful formula which is used in the main
context.
 
{\bf Note added}: While writing this manuscript, there appeared a
paper \cite{sug029} which treats the same subject. However, the action
obtained in that paper seems to be non-supersymmetric while the
background geometry is obviously supersymmetric.

\section{Supersymmetry of IIA pp-wave geometry}

The IIA pp-wave geometry Eq.~(\ref{pp-wave}) has 24 Killing spinors.
The original eleven-dimensional pp-wave background is maximally
supersymmetric and therefore has 32 Killing spinors.  Among them only
24 components survive after the compactification along $x^9$. This can
be shown by counting the number of spinors which are invariant under
the Lie derivative along $x^9$-direction.\footnote{See
Ref. \cite{mic140} for detailed derivation.}

Here we directly compute the ten-dimensional Killing spinor equations
and get the explicit expression for those Killing spinors. This will
be needed in the study of supersymmetry in the IIA string theory and
also it will illuminate the nature of supersymmetry more clearly.  The
Killing spinor equations come from the vanishing of supersymmetry
variations of gravitino and dilatino fields,
$\delta_\eta\psi_\mu=\delta_\eta\lambda=0$. The general expressions for these
supersymmetry transformation rules are given in the appendix.

The equation from the dilatino transformation rule, $\delta_\eta\lambda=0$, 
reduces to the condition
\begin{equation}
\Gamma^+(\Gamma^{12349}-1)\eta =0~.
\label{cond1}
\end{equation} 
From the gravitino transformation rule, we have another condition
which is
\begin{eqnarray}
\delta_\eta \psi_\mu =  (\nabla_\mu + \Omega_\mu ) \eta = 0~,  
\label{cond2}
\end{eqnarray} 
where the explicit expression for $\Omega_\mu$ is given in the appendix. 
In our case at hand, $\Omega_{\pm}$ are 
\begin{eqnarray}
\Omega_- &=& 0~, \nonumber \\
\Omega_{+} &=&-
\frac{\mu}{96}\Gamma^{123}\left(\Gamma^{-+}(9-\Gamma^{12349}) +
  (15-7\Gamma^{12349})\right)~.  
\end{eqnarray}  
For the spinors $\eta$ which satisfy Eq.~(\ref{cond1}), $\Omega_I$'s
are given by
\begin{eqnarray}
\Omega_i &=& -\frac{\mu}{6}\Gamma^{123}\Gamma^+\Gamma^i~, 
                                        \nonumber \\
\Omega_{i^\prime} &=&
-\frac{\mu}{12}\Gamma^{123}\Gamma^+\Gamma^{i^\prime}~.
\label{omegaI}
\end{eqnarray}  

In order to see the structure of the spinor more clearly and to solve
the Killing spinor equations explicitly, we now introduce the
following representations for $SO(1,9)$ gamma matrices:
\[
 \Gamma^0= -i\sigma^2 \otimes {\bf 1}_{16}~,~~~
 \Gamma^{11}= \sigma^1 \otimes {\bf 1}_{16}~,~~~
 \Gamma^I= \sigma^3 \otimes \gamma^I~,
\]
\begin{equation}
 \Gamma^9= -\sigma^3 \otimes \gamma^9~,~~~
 \Gamma^{\pm} = \frac{1}{\sqrt{2}}(\Gamma^0 \pm \Gamma^{11})~,
\label{gamma}
\end{equation}
where $\sigma$'s are Pauli matrices, and ${\bf 1}_{16}$ the $16 \times
16$ unit matrix. $\gamma^I$ are the $16 \times 16$ symmetric real
gamma matrices satisfying the spin$(8)$ Clifford algebra $\{ \gamma^I,
\gamma^J \} = 2 \delta^{IJ}$, which are reducible to the ${\bf
8_s}+{\bf 8_c}$ representation of spin$(8)$.  We note that, since we
compactify along the $x^9$ direction, $\Gamma^9$ is the $SO(1,9)$
chirality operator and $\gamma^9$ becomes $SO(8)$ chirality operator,
\[
\gamma^9 = \gamma^1 \cdots \gamma^8~.
\]

Firstly, we consider the 16 component spinor  which satisfies
\begin{equation}
\Gamma^+ \tilde{\eta}=0~, 
\end{equation}
which is clearly a solution of the condition, Eq.~(\ref{cond1}).  With
the above gamma matrix representation, Eq.~(\ref{gamma}), the
transformation parameter $\tilde{\eta}$ is of the form
\begin{equation}
\tilde{\eta} = \left( 
              \begin{array}{c} 0 \\ \tilde{\epsilon} \end{array}
       \right) ~.  \label{kin}
\end{equation}
By plugging this into the condition Eq.~(\ref{cond2}), we see from the
$-$ and $I$ components of the condition that $\tilde{\eta}$ is
independent of the coordinates $x^-$ and $x^I$, and hence
$\tilde{\epsilon}$ may be at most a function of $x^+$.

The $+$ component of Eq.~(\ref{cond2}), which is the only remaining
nontrivial condition and specifies the light-cone time dependence of
the spinor $\tilde{\epsilon}$, is read as
\begin{equation}
\partial_+ \tilde{\epsilon}
 = -\frac{\mu}{4}\gamma^{123}
   (1-\frac{1}{3}\gamma^{12349})\tilde{\epsilon}~.
\label{+cond}
\end{equation}
Since $(\gamma^{12349})^2=1$, it is now convenient to decompose
$\tilde{\epsilon}$ as $\tilde{\epsilon}=\tilde{\epsilon}^+ +
\tilde{\epsilon}^-$ where $\tilde{\epsilon}^\pm$ are eigenstates of
$\gamma^{12349}$:
\begin{equation}
\gamma_{12349}\tilde{\epsilon}^{\pm} 
 = \pm \tilde{\epsilon}^{\pm}~.
\end{equation}
Then solving the Eq.~(\ref{+cond}) for each of $\tilde{\epsilon}^\pm$
leads to
\begin{eqnarray}
\tilde{\epsilon}^+ 
  &=& e^{-\frac{\mu}{6}\gamma^{123}x^+} \tilde{\epsilon}_0^+~, 
                                        \nonumber \\
\tilde{\epsilon}^- 
  &=& e^{-\frac{\mu}{3}\gamma^{123}x^+} \tilde{\epsilon}_0^-~,
\end{eqnarray}
where $\tilde{\epsilon}_0^{\pm}$ are constant spinors with eight
independent components. Therefore the 16 component Killing spinor
$\tilde{\eta}$ considered so far depends only on $x^+$ and, as will be
shown in the next section, correspond to 16 kinematical supersymmetry
of the IIA string action.

Next we consider eight remaining Killing spinors, which will
correspond to the dynamical supersymmetry of the string theory.  First
of all, from the condition, Eq.~(\ref{cond1}), the transformation
parameter of the dynamical supersymmetry can be written of the form,
\begin{equation}
\eta = \left( \begin{array}{c} \epsilon \\ \epsilon' \end{array}
       \right) ~,
 \label{dyn}
\end{equation} 
where $\epsilon$ should satisfy $\gamma^{12349}\epsilon=-\epsilon$.
The $x^-$ component of the Killing spinor equation (\ref{cond2})
reduces to $\partial_-\eta =0$, which means that $\eta$ is independent
of $x^-$.  The $x^I$ components of Eq.~(\ref{cond2}) reduce to
$(\partial_I + \Omega_I ) \eta =0$. Since $\Omega_I\Omega_J=0$ for any
$I, J =1, \cdots , 8$ due to the fact that $(\Gamma^+)^2=0$, $\eta$
depends on $x^I$, at most, linearly and thus is of the following form
\begin{eqnarray}
\eta = (1+x^I\Omega_I) \left( \begin{array}{c} \epsilon \\ 0
    \end{array} \right)~,
\label{epsilon}
\end{eqnarray}
where $\epsilon$ is independent of $x^I$ as well as $x^-$. One can
easily convince that this automatically satisfies $(\nabla_+ +
\Omega_+ ) \eta =0$ as well only if $\epsilon$ is independent of
$x^+$. This means that the eight dynamical supersymmetry in the
light-cone IIA superstring action, which corresponds to this Killing
spinor, is independent of time and thus becomes genuine symmetry which
commute with the Hamiltonian. The fermions in the matrix model in
eleven dimensional pp-wave background have different masses from those
of bosons as the supersymmetry in matrix model does not commute with
the Hamiltonian. Now it is natural to expect to get the same mass
content in bosons and fermions, which is indeed the case as will be
shown in the next section.

\section{Light-cone IIA superstring action in plane wave background}

The best way to get the IIA GS superstring action in the general
background is the double dimensional reduction of the supermembrane
action in eleven-dimensions\cite{duf70}.  In general, it is very
complicate to get the full expression of the GS superstring action in
the general background, which is up to 32th order in terms of the
fermionic coordinate. However, in the case at hand, we know the full
action of supermembrane in eleven-dimensional pp-wave background. This
comes from the fact that eleven-dimensional pp-wave geometry can be
thought as a special limit \cite{pen271,bla081} of $AdS_4 \times S^7$
geometry on which the full supermembrane action is constructed using
coset method \cite{dew209}. Therefore we start from the supermembrane
action in eleven-dimensional pp-wave in the rotated coordinates
Eq.~(\ref{pp-wave-2}), given in the appendix. The supermembrane is
wrapped on $x^9$, and becomes IIA superstring after the double
dimensional reduction along $x^9$.

The general expression of the eleven dimensional supermembrane action
is too complicated, especially for the double dimensional
reduction. Thus we first simplify the action by fixing the fermionic
$\kappa$-symmetry,
\begin{equation}
\Gamma^+ \theta = 0~.
\label{kfix}
\end{equation}
Super elfbein of Ref.~\cite{dew209} in this fixing condition has the
following form:\footnote{The index notations adopted here are as
follows: $M,N,...$ are used for the target superspace indices while
$A,B,...$ for tangent superspace. As usual, a superspace index is the
composition of two types of indices such as $M=(\mu,\alpha)$ and
$A=(r,a)$. $\mu,\nu,...~(r,s,...)$ are the ten dimensional target
(tangent) space-time indices.  $\alpha,\beta,...~(a,b,...)$ are the
ten dimensional (tangent) spinor indices.  For the eleven dimensional
case, we denote quantities and indices with hat to distinguish from
those of ten dimensions.  $m,n,...$ are the worldsheet vector indices
with values $\tau$ and $\sigma$. The convention for the worldsheet
antisymmetric tensor is taken to be $\epsilon^{\tau\sigma}=1$.}
\begin{eqnarray}
\hat{E}^a &=& D \theta ~,   \nonumber \\
\hat{E}^{\hat{r}} &=& \hat{e}^{\hat{r}} 
        + \bar{\theta} \Gamma^{\hat{r}} D \theta~.
\end{eqnarray}
where $\bar{\theta} = i \theta^T \Gamma^0$ and $D \theta$ is the
super-covariant one-form whose general expression is given by
\begin{equation}
D \theta = d \theta 
        + \frac{1}{4} \hat{\omega}^{\hat{r}\hat{s}}
                \Gamma_{\hat{r}\hat{s}} \theta
        + \frac{1}{2! 3! 4!} \hat{e}^{\hat{r}} 
        ( \Gamma_{\hat{r}}{}^{\hat{s}\hat{t}\hat{u}\hat{v}}
         -8 \delta_{\hat{r}}^{[ \hat{s}}
            \Gamma^{\hat{t}\hat{u}\hat{v}]} ) \theta
        F_{\hat{s}\hat{t}\hat{u}\hat{v}}~.
\end{equation}
In the pp-wave background geometry, Eq.~(\ref{pp-wave-2}), and in the
$\kappa$-symmetry fixing condition, Eq.~(\ref{kfix}), $D\theta$
reduces to
\begin{equation}
D \theta = d \theta 
  - \frac{\mu}{4} 
    \left( \Gamma^{123} + \frac{1}{3} \Gamma^{49} \right)
    \theta dx^+ ~,
\end{equation}
where $\Gamma^{123}$ comes from the non-vanishing constant four-form
field strength and $\Gamma^{49}$ is due to the component of the spin
connection $\hat{\omega}^{49}_+ = -\mu / 6$, which is related to the
ten dimensional RR two-form field strength as $\hat{\omega}^{49}_+ =
F_{+4}/2$ under the Kaluza-Klein reduction.  In the $\kappa$-symmetry
fixing condition, the super three-form field is simplified as
\begin{equation}
\hat{B}
   = \frac{1}{2} \hat{e}^+ \wedge \hat{e}^i \wedge \hat{e}^j
        \hat{C}_{+ij}
   - \bar{\theta} \Gamma_{+\hat{I}} D \theta \wedge \hat{e}^+
        \wedge \hat{e}^{\hat I} ~,
\label{3f}
\end{equation}
where ${\hat I} = 1,...,9$.

Through the usual Kaluza-Klein reduction, the super zehnbein is
related to the above super elfbein.  Among the super zehnbein fields,
$E^r$ is the only what we need for the construction of the superstring
action and is obtained from the relations $\hat{E}^r_\mu = \Phi^{-1/3}
E^r_\mu$ and $\hat{E}^r_\alpha = \Phi^{-1/3} e^{\phi/6} E^r_\alpha$
where $\Phi$ is the super dilaton field given by
$\hat{E}^9_9$ \cite{cve202}.  Their explicit expression in the 
component form is then
\begin{eqnarray}
E^r_\mu &=& e^r_\mu - \frac{\mu}{4} \delta^r_- \delta_\mu^+ 
        \bar{\theta} \Gamma^- \left( \Gamma^{123} 
          + \frac{1}{3} \Gamma^{49} \right) \theta ~,
                                \nonumber \\
E^r_\alpha &=& - (\bar{\theta} \Gamma^r)_\alpha ~.
\label{10bein}
\end{eqnarray}

Having the component form of the ten dimensional super fields, it is
ready to construct the Type IIA superstring action in the pp-wave
background, Eq.~(\ref{pp-wave}).  In the superspace formalism, the IIA
superstring action is written as follows \cite{duf70}:
\begin{equation}
S_{IIA} = \frac{1}{2\pi \alpha'} \int d^2 \sigma
 \left( - \frac{1}{2} \sqrt{-h} h^{mn} \Pi_m^r \Pi_n^s \eta_{rs}
        + \frac{1}{2} \epsilon^{mn} \Pi_m^A \Pi_n^B B_{BA}
 \right)~,
\label{superIIA}
\end{equation}
where $\Pi^r_m$ is the pullback of super zehnbein onto the worldsheet;
\[
\Pi^r_m = \partial_m Z^M E^r_M~.
\]
By using the super zehnbein fields, Eq.~(\ref{10bein}), each
component of the pullback in the background, Eq.~(\ref{pp-wave}), is
expressed as
\begin{eqnarray}
\Pi^+_m &=& \partial_m X^+ ~,   \nonumber \\
\Pi^-_m &=& \partial_m X^- 
      + \partial_m X^+ 
      \left[ \frac{1}{2} A(X^I) - \frac{\mu}{4}
              \bar{\theta} \Gamma^- 
                 \left( \Gamma^{123} + \frac{1}{3}
               \Gamma^{49} \right) \theta 
      \right] 
      + \bar{\theta} \Gamma^- \partial_m \theta ~,
                           \nonumber \\
\Pi^I_m &=& \partial_m X^I ~.
\label{pullb}
\end{eqnarray}
On the other hand, the super three-form field, Eq.~(\ref{3f}), leads
to the following Wess-Zumino term:
\begin{eqnarray}
\frac{1}{2} \epsilon^{mn} \Pi^A_m \Pi^B_n B_{BA}
  &=& \frac{1}{2} \epsilon^{mn} \partial_m Z^M \partial_n Z^N
      \hat{B}_{9NM}      \nonumber \\
  &=& - \epsilon^{mn} \partial_m X^+ 
       (\bar{\theta} \Gamma^{-9} \partial_n \theta)~.
\label{wzt}
\end{eqnarray}

Plugging the expressions, Eqs.~(\ref{pullb}) and (\ref{wzt}) into the
super field action, Eq.~(\ref{superIIA}), we then finally get the
$\kappa$-symmetry fixed action in the component form for the Type IIA
superstring in the pp-wave background, Eq.~(\ref{pp-wave}), which is
obtained as
\begin{eqnarray}
S_{IIA} 
  &=& -\frac{1}{4 \pi \alpha'} \int d^2 \sigma \sqrt{-h} h^{mn}
  \bigg[ -2 \partial_m X^+ \partial_n X^- 
    + \partial_m X^I \partial_n X^I 
    - A(X^I) \partial_m X^+ \partial_n X^+
                         \nonumber \\
  & & -2 \partial_m X^+ \bar{\theta} \Gamma^- \partial_n \theta
      + \frac{\mu}{2} \partial_m X^+ \partial_n X^+
       \bar{\theta} \Gamma^- 
         \left( \Gamma^{123} + \frac{1}{3} \Gamma^{49}
         \right) \theta
  \bigg]    
                         \nonumber \\
  & &  - \frac{1}{2 \pi \alpha'} \int d^2 \sigma
   \epsilon^{mn} \partial_m X^+ \bar{\theta} \Gamma^{-9} 
    \partial_n \theta ~.
\label{kfix-action}
\end{eqnarray}
  
We now consider the bosonic light-cone gauge to get the action for the
physical degrees of freedom.  The equation of motion for $X^+$ is
harmonic, the same as in the flat case, which means that the usual
light-cone gauge, $X^+ \propto \tau$, is allowed.  We then take the
conventional light-cone gauge conditions as follows:
\begin{eqnarray}
X^+  &=& \alpha' p^+ \tau ~,   \nonumber \\
\sqrt{-h} &=& 1 ~,   \nonumber \\
h_{\sigma \tau} &=& 0 ~.
\label{lcg}
\end{eqnarray}
where $p^+$ is the total momentum congugate to $X^-$.  The last two
conditions are for the fixing of the worldsheet diffeomorphisms and
allow us to fix other worldsheet metric components consistently as
\[
- h_{\tau \tau} = h_{\sigma \sigma} = 1 ~.
\]
In this bosonic light-cone gauge choice, the superstring action,
Eq.~(\ref{kfix-action}), is further simplified, and the light-cone
string action, which we call $S_{LC}$, is read as
\begin{eqnarray}
S_{LC} 
 &=& - \frac{1}{4 \pi \alpha'} \int  d^2 \sigma
 \Bigg[ \eta^{mn} \partial_m X^I \partial_n X^I 
      + \frac{m^2}{9} (X^i)^2
      + \frac{m^2}{36} (X^{i'})^2
                       \nonumber \\
 & & + \bar{\theta} \Gamma^- \partial_\tau \theta 
     + \bar{\theta} \Gamma^{-9} \partial_\sigma \theta 
     - \frac{m}{4} \bar{\theta} \Gamma^- 
        \left( \Gamma^{123} + \frac{1}{3} \Gamma^{49} \right)
        \theta  
  \Bigg] ~,
\end{eqnarray}
where we have rescaled the fermionic coordinate as $\theta \rightarrow
\theta / \sqrt{2 \alpha' p^+}$.  $m$ is a mass parameter defined by
\[
 m \equiv \mu \alpha' p^+ ~,
\]
which characterizes the masses of the worldsheet fields.  We see that
the light-cone gauge fixed action $S_{LC}$ is quadratic in fields and
thus describes a free theory as in the IIB case \cite{met044}. 

Though it is now obvious that $S_{LC}$ describes the theory of free
fields, the mass spectrum is still unclear due to the mass term of
fermionic fields.  Reading off the mass contents of the fermionic
fields is necessary and important step, especially to see the
supersymmetry of the action more clearly, which will be elucidated in
the next section.  We first rewrite the action in the 16 component
spinor notation with $\theta^A = \frac{1}{2^{1/4}} \left(
\begin{array}{c} 0 \\ \psi^A
\end{array} \right)$ (Superscript $A$ denotes the $SO(1,9)$
chirality.), under which we have
\begin{eqnarray}
S_{LC}
 &=&  - \frac{1}{4 \pi \alpha'} \int  d^2 \sigma
 \Bigg[ \eta^{mn} \partial_m X^I \partial_n X^I 
      + \frac{m^2}{9} (X^i)^2
      + \frac{m^2}{36} (X^{i'})^2
                       \nonumber \\
 & & - i \psi^1 (\partial_\tau + \partial_\sigma ) \psi^1
     - i \psi^2 (\partial_\tau - \partial_\sigma ) \psi^2
     - i \frac{m}{2} \psi^2 \left( \gamma^{123} 
                + \frac{1}{3} \gamma^4 \right) \psi^1
 \Bigg]~.
\label{lc-action}
\end{eqnarray}
Let us observe that $(\gamma^{1234})^2=1$ and $ [ \gamma^{1234},
\gamma^9 ] = 0$, which mean that a fermionic field with definite
$SO(8)$ chirality can be decomposed according to the eigenvalues of
$\gamma^{1234}$: $\psi^A = \psi^A_+ + \psi^A_-$ where $\psi^A_\pm$ are
eigenstates of $\gamma^{1234}$, that is, $\gamma^{1234} \psi^A_\pm =
\pm \psi^A_\pm$.  With this decomposition, the fermion mass term in
the light-cone action, Eq.~(\ref{lc-action}), can be rewritten as
\begin{equation}
-2i \frac{m}{3} \psi^2_+ \gamma^4 \psi^1_-
+2i \frac{m}{6} \psi^2_- \gamma^4 \psi^1_+ ~.
\end{equation}
In our notation, fermion has the same $SO(1,9)$
and $SO(8)$ chirality measured by $\Gamma^9$ and $\gamma^9$,
respectively.  Thus, among sixteen fermionic components in total,
eight with $\gamma^{12349}=1$ have the mass of $m/6$ and the other
eight with $\gamma^{12349}=-1$ the mass of $m/3$, which are identical
with the masses of bosons. Therefore the theory contains two
supermultiplets $(X^i, \psi^1_-, \psi^2_+)$ and $(X^{i'}, \psi^1_+,
\psi^2_-)$ of (4,4) supersymmetry with the masses $m/3$ and $m/6$,
respectively.

\section{${\cal N}$=(4,4) Worldsheet Supersymmetry}

In this section we describe the supersymmetry in the above light-cone
fixed action, Eq.~(\ref{lc-action}).  The supersymmetry
transformation $(\delta_\eta)$ for the worldsheet fields is the odd
part of the supertranslation in superspace, which is read off from
\begin{eqnarray}
\delta_\eta Z^M E_M^r &=& 2 \bar{\theta} \Gamma^r \eta~, \nonumber \\
\delta_\eta Z^M E_M^a &=& \eta^a~.
\end{eqnarray}
With these relations, the supersymmetry transformation rules in the
light-cone gauge, Eq.~(\ref{lcg}), and the $\kappa$ symmetry fixing
condition $\Gamma^+ \theta = 0$ are obtained as
\begin{eqnarray}
\delta_\eta X^+ &=& 0~, \nonumber \\
\delta_\eta X^- &=& \bar{\theta}^A \Gamma^- \eta^A~, \nonumber \\
\delta_\eta X^I &=& \bar{\theta}^A \Gamma^I \eta^A~, \nonumber \\
\delta_\eta \theta^A &=& \eta^A~.
\end{eqnarray}

For the kinematic supersymmetry $\delta_{\tilde{\eta}}$ with the
parameter $\tilde{\eta}$ of Eq.~(\ref{kin}) satisfying $\Gamma^+
\tilde{\eta} = 0$, the light-cone gauge choice $X^+ = \alpha' p^+
\tau$ is preserved and the transformation rules for the physical
degrees of freedom are
\begin{eqnarray}
\tilde{\delta} X^I &=& 0~,   \nonumber \\
\tilde{\delta} \theta^A &=& \tilde{\eta}^A~.
\end{eqnarray}
These rules tell us that the kinematic supersymmetry is nonlinearly
realized on the string worldsheet.

For the dynamical supersymmetry $\delta_\eta$ with the parameter of
Eq.~(\ref{epsilon}), fermionic $\kappa$-symmetry fixing condition is
no longer preserved, because $\Gamma^+ \eta \neq 0$. Therefore we need
the supplement $\kappa$ transformations so that the total
transformation rule,
\[
\delta = \delta_\eta + \delta_\kappa
\]
preserves $\kappa$-symmetry fixing condition.
In the superspace notation, the $\kappa$ symmetry transformation
$(\delta_\kappa)$ satisfies the following equations
\begin{eqnarray}
\delta_\kappa Z^M E_M^r &=& 0~, \nonumber \\
\delta_\kappa Z^M E_M^a &=& (1-\Gamma \Gamma^9)^a{}_b \kappa^b~,
\end{eqnarray}
where the matrix $\Gamma$ is given by
\begin{equation}
\Gamma = \frac{1}{2 \sqrt{-g}} \epsilon^{mn} 
     \Pi_m^r \Pi_n^s \Gamma_r \Gamma_s
\label{gdef}
\end{equation}
with the determinant $g$ of the induced metric $g_{mn}$ given by
$g_{mn}=\Pi_m^r \Pi_n^s \eta_{rs}$.  $\Gamma$ has the properties as a
projection operator:
\begin{equation}
\Gamma^2 = 1 ~,~~~{\rm Tr} \Gamma = 0 ~.
\label{gprop}
\end{equation}
In the component notation, the transformation rules for the
$\kappa$-symmetry in our gauge choice become
\begin{eqnarray}
\delta_\kappa X^+ &=& 0 ~,   \nonumber \\
\delta_\kappa X^- 
  &=& \delta_\kappa \bar{\theta}^A \Gamma^- \theta^A ~,
                           \nonumber \\
\delta_\kappa X^I 
  &=&  \delta_\kappa \bar{\theta}^A \Gamma^I \theta^A ~,
                           \nonumber \\
\delta_\kappa \theta^A &=& (1-\Gamma \Gamma^9) \kappa^A ~.
\end{eqnarray}

For the superstring case, it is usually convenient to introduce
\begin{equation}
\kappa^A_m = -i \frac{\sqrt{-h}}{2 \sqrt{-g}} \Pi_m \Gamma \kappa^A ~,
\end{equation}
which allows us to view the parameter $\kappa$ as a worldsheet vector.
It is easily checked using Eqs.~(\ref{gdef}) and (\ref{gprop}) that
$\kappa^2_m$ $(\kappa^1_m)$ is the (anti-) self-dual vector and thus
each worldsheet vectors has one independent component, say $\rho^A$.
In our notation, $\kappa^{1\tau}=-\kappa^{1\sigma} = -\rho^1$ and
$\kappa^{2\tau}=\kappa^{2\sigma} = -\rho^2$.  Then the $\delta_\kappa
\theta^A$ becomes
\begin{equation}
\delta_\kappa \theta^A = 2 i \Pi_m^r \Gamma_r \kappa^{Am}~.
\end{equation}

Now one can find the appropriate $\kappa$ transformation parameters so
that the total transformation rules obey $\Gamma^+\delta
\theta=0$.\footnote{For example, see Ref. \cite{hyu247}.} The
resultant transformation rules are as follows:
\begin{eqnarray}
\delta X^I &=& \frac{i}{\sqrt{\alpha' p^+}} 
                \psi^A \gamma^I \epsilon^A  ~,   \nonumber \\
\delta \psi^1 &=&
  \frac{1}{\sqrt{\alpha' p^+}} 
        (\partial_\tau X^I-\partial_\sigma X^I) 
        \gamma^I \epsilon^1
 +\frac{1}{\sqrt{\alpha' p^+}} 
        \left( \frac{m}{3} X^i \gamma^{123} \gamma^i 
              +\frac{m}{6} X^{i'} \gamma^{123} \gamma^{i'}
        \right) \epsilon^2 ~,
                                                \nonumber \\
\delta \psi^2 &=&
  \frac{1}{\sqrt{\alpha' p^+}} 
        (\partial_\tau X^I + \partial_\sigma X^I) 
        \gamma^I \epsilon^2
 +\frac{1}{\sqrt{\alpha' p^+}} 
        \left( \frac{m}{3} X^i \gamma^{123} \gamma^i 
              +\frac{m}{6} X^{i'} \gamma^{123} \gamma^{i'}
        \right) \epsilon^1 ~,
\label{susy}
\end{eqnarray}                  
where we performed the rescaling in fermions $\psi \rightarrow \psi /
(2^{1/4} \sqrt{2 \alpha' p^+})$ and the supersymmetry parameter
$\epsilon \rightarrow \epsilon / 2^{1/4}$ given in
(\ref{epsilon}). Now it is straightforward to check that the
light-cone action Eq.~(\ref{lc-action}) is on-shell invariant under the
above (4,4) supersymmetry, Eq.~(\ref{susy}).

\section{Discussions}

The simplicity of the action of the IIA GS superstring on the pp-wave
geometry and its supersymmetry is quite striking. The light-cone
action consists only of quadratic terms and thus the theory is
free. Furthermore the dynamical supersymmetry is independent of the
worldsheet time coordinate which implies that the fields in the same
supermultiplet have the same mass.  Since the number of dynamical
supersymmetries is half the number of those in IIA superstring on the
flat background, the full field contents split into two
supermultiplets. Each supermultiplet consists of half number of bosons
and fermions which has the same mass. Therefore the theory can be
interpreted as the free $\cal{N}$=(4,4) two-dimensional theory with
two massive supermultiplets.

The mode expansions are straightforward and the spectrum should have
the same supersymmetry structure.  It should be much more
straightforward to find BPS D-branes in the string theory
\cite{dab231,ske054,kns025,ak134,bis042} than the original
eleven-dimensional M theory. For example, all the BPS states found in
\cite{hyu090,bak033,ali237} in the context of matrix model in eleven
dimensions would be found in this string theory.  The IIA matrix
string theory on pp-wave background should also have the same
supermultiplet structure as it becomes free string theory in the IR
limit.  In that sense, the matrix string theory on ten dimensional
pp-wave seems to be more transparent than the matrix model on eleven
dimensional pp-wave.  All these will be presented in the forthcoming
paper \cite{hyuxxx}.

\vspace{12mm} \noindent {\Large \bf Appendix}

\appendix\setcounter{section}{0} 
In this appendix, we describe how to get the ten-dimensional IIA
pp-wave through the dimensional reduction from the eleven-dimensional
pp-wave. We also give some useful formula used in our computations.
 
The pp-wave geometry in eleven-dimensions  is given by
\begin{eqnarray}
& & ds_{11}^2 = - 2 dX^+ dX^-
    - \left( \sum^3_{i=1} \frac{\mu^2}{9} (X^i)^2
            +\sum^9_{i'=4} \frac{\mu^2}{36} (X^{i'})^2
      \right) (dX^+)^2
    + \sum^9_{I=1} (dX^I)^2~,
                                      \nonumber \\
& & F_{+123} = \mu~,
\label{pp-wave-1}
\end{eqnarray}
where $\mu$ is a parameter of the geometry and characterizes the
matrix theory of Ref.~\cite{ber021}. 

We change the coordinates as 
\begin{eqnarray}
X^+= x^+~, \ \ \ X^- = x^- - \frac{\mu}{6}x^4 x^9~, \ \ \ X^I = x^I~,
\ \ \ I= 1, 2, 3, 5 \cdots 8~, \cr 
X^4 = x^4 \cos ( \frac{\mu}{6}x^+)
- X^9 \sin ( \frac{\mu}{6}X^+)~, \ \ \ 
X^9 = x^4 \sin (\frac{\mu}{6}x^+) + x^9 \cos ( \frac{\mu}{6}x^+)~,
\end{eqnarray}
under which the geometry becomes
\begin{eqnarray}
& & ds_{11}^2 = - 2 dx^+ dx^-
    - \left( \sum^3_{i=1} \frac{\mu^2}{9} (x^i)^2
            +\sum^8_{i'=5} \frac{\mu^2}{36} (x^{i'})^2
      \right) (dx^+)^2 + \frac{2}{3}\mu x^4 dx^9dx^+
    + \sum^9_{I=1} (dx^I)^2~,
                                      \nonumber \\
& & F_{+123} = \mu~.
\label{pp-wave-2}
\end{eqnarray}
Therefore in the new coordinates, $\partial_9=\frac{\partial}{\partial
x^9}$ is an isometry.  If the $x^9$ coordinate is periodically
identified, M theory in this pp-wave background can be compactified to
give type IIA string theory in the following new pp-wave geometry:
\begin{eqnarray}
& & ds^2 = - 2 dx^+ dx^-
    - A(x^I) (dx^+)^2  + \sum^8_{I=1} (dx^I)^2~,
                                      \nonumber \\
& & F_{+123} = \mu~,\ \ \  F_{+4} = -\frac{\mu}{3} ~,
\end{eqnarray}
where
\begin{equation}
A(x^I) = \sum^4_{i=1} \frac{\mu^2}{9} (x^i)^2
            +\sum^8_{i'=5} \frac{\mu^2}{36} (x^{i'})^2 ~.
\end{equation}

From this ten dimensional IIA geometry, we may choose the zehnbein as
\begin{eqnarray}
e^+ &=& dx^+ ~, \nonumber \\
e^- &=& dx^- + \frac{1}{2} A(x^I) dx^+ ~, \nonumber \\
e^I &=& dx^I~.
\end{eqnarray}
The only non-vanishing spin connection one-form is 
\begin{equation}
\omega^{-I}= \frac{1}{2}\partial_I A(x^I) dx^+ ~.
\end{equation}

The supersymmetry transformation rule for the ten-dimensional
gravitino $\psi_\mu$, in the string frame and in the vanishing fermion
background, is
\begin{eqnarray}
\delta_\eta \psi_\mu =  (\nabla_\mu + \Omega_\mu ) \eta~,  
\end{eqnarray} 
where the covariant derivative is given by
$$\nabla_\mu = \partial_\mu + \frac{1}{4} \omega_\mu^{~rs}
\Gamma_{rs}$$ and 
\begin{eqnarray}
\Omega_\mu 
 &=&
   -\frac{1}{8} \Gamma_r \Gamma_s \eta
        e_\mu^{~r} e^{\nu s} \partial_\nu \phi
  -\frac{1}{64} e_\mu^{~r} ( \Gamma_{rst}- 14 \eta_{rs} \Gamma_t)
     \Gamma^{11} \eta e^\phi F^{st}
                                                \nonumber \\
 & & +\frac{1}{96} e_\mu^{~r}
     (\Gamma_r^{~stu}-9 \delta_r^s \Gamma^{tu}) \Gamma^{11} \eta
      H_{stu}
     +\frac{1}{768} e_\mu^{~r}
     ( 3\Gamma_r^{~stuv}-20 \delta_r^s \Gamma^{tuv} ) \eta
     e^\phi F'_{stuv} ~.
\label{dpsi}
\end{eqnarray}
The dilatino field transforms under the supersymmetry as
\begin{eqnarray}
\delta_\eta \lambda
 &=& -\frac{1}{2 \sqrt{2} } \Gamma^r \Gamma^{11} \eta
            e_r^{~\nu} \partial_\nu \phi
    -\frac{3}{16 \sqrt{2} } \Gamma^{rs} \eta e^\phi F_{rs}
                                               \nonumber \\
 & & + \frac{1}{24 \sqrt{2}} \Gamma^{rst} \eta H_{rst}
    + \frac{1}{192 \sqrt{2} } \Gamma^{rstu} \Gamma^{11}
            \eta  e^\phi F'_{rstu} ~.
                                               \label{dlam}
\end{eqnarray}

\section*{Acknowledgments}
One of us (H.S.) would like to thank the Yonsei Visiting Research
Center (YVRC) for its hospitality, where this work has been completed.
The work of S.H. was supported in part by grant No. R01-2000-00021
from the Basic Research Program of the Korea Science and Engineering
Foundation.

\newpage

\end{document}